# Neuromorphic computing using wavelength-division multiplexing

Xingyuan Xu, Weiwei Han, Mengxi Tan, Yang Sun, Yang Li, Jiayang Wu, Roberto Morandotti, *Fellow IEEE*, Arnan Mitchell, *Senior Member IEEE*, Kun Xu, and David J. Moss, *Fellow IEEE*

*Abstract*—**Optical neural networks (ONNs), or optical neuromorphic hardware accelerators, have the potential to dramatically enhance the computing power and energy efficiency of mainstream electronic processors, due to their ultra-large bandwidths of up to 10's of terahertz together with their analog architecture that avoids the need for reading and writing data back-and-forth. Different multiplexing techniques have been employed to demonstrate ONNs, amongst which wavelength-division multiplexing (WDM) techniques make sufficient use of the unique advantages of optics in terms of broad bandwidths. Here, we review recent advances in WDM-based ONNs, focusing on methods that use integrated microcombs to implement ONNs. We present results for human image processing using an optical convolution accelerator operating at 11 Tera operations per second. The open challenges and limitations of ONNs that need to be addressed for future applications are also discussed.**

## I. INTRODUCTION

Artificial neural networks (ANNs), inspired by the human biological brain, have achieved unprecedented success in a wide range of applications ranging from image recognition to sophisticated board games [1-10], due to their capability for learning to be able to process unknown data intelligently. ANNs are mathematical network models formed by densely interconnected neurons, with the ability to address complicated tasks, limited only by the scale of the network (i.e., number of neurons and synapses). As such, the number of operations and parameters of ANNs, which determine the

Manuscript received XXX X, XXXX; revised XXX X, XXXX; accepted XXX X, XXXX. Date of publication XXX X, XXXX; date of current version XXX X, XXXX. This work was supported by the National Natural Science Foundation of China (NSFC) under Grant 62135009, the ECR-SUPRA program, the Beijing Natural Science Foundation (No. Z180007). *(Corresponding author: David J. Moss)*

X. Xu, W. Han and K. Xu are with the State Key Laboratory of Information Photonics and Optical Communications, Beijing University of Posts and Telecommunications, Beijing 100876, China.

Y. Sun, Y. Li, J. Wu, and D. J. Moss are with Optical Sciences Center, Swinburne University of Technology, Hawthorn, VIC 3122, Australia. (e-mail: dmoss@swin.edu.au)

M. Tan and A. Mitchell are with RMIT University, Melbourne, VIC 3001, Australia.

R. Morandotti is with the INRS-Énergie, Matériaux et Télécommunications, 1650 Boulevard Lionel-Boulet, Varennes, Québec, J3X 1S2, Canada.

Color reproduction of one or more of the figures in this letter are available online at http://ieeexplore.ieee.org.

Digital Object Identifier 

hardware's computing power, scale exponentially with performance, including key attributes such as accuracy [11].

However, while more advanced applications of ANNs bring about ever-higher demands of the hardware's computing capabilities, the performance density and energy efficiency of leading electronic hardware platforms face severe limitations, as reflected by Moore's law [11-14]. The performance density (i.e., number of operations performed within a given chip scale) can no longer increase beyond where Moore's law ends, at device feature sizes of around 5 nm. As a result of reaching this limit, the energy efficiency of memories has shown little improvement since 2015. This is due to two fundamental limitations – the so-called electronic bandwidth bottleneck and the von-Neumann bottleneck [15, 16]: the former limits the clock rate of electronic devices to ~ 2 GHz, while the latter introduces high energy consumption during the process of reading and writing data back and forth.

Optical neural networks (ONNs), are promising next-generation neuromorphic accelerators for ANNs, since they can potentially offer ultra-large bandwidths of >30 THz in order to reach dramatically accelerated computing speeds, together with low power consumption due to operating inherently in the analog regime [17-64]. To realize ONNs, weighted synapses forming the interconnections between neurons need to be implemented with multiple physical paths established within the temporal-, spatial- and wavelength-division parallelism [17-25]. For example, time-delayed optical loops are employed to achieve reservoir computing or spiking neural networks [26-35] that can store over one thousand nodes within the cavity. Integrated photonic waveguides [36-39] and free-space optics [40-45] have been employed in order to achieve synaptic connections using multiple wavelengths. Further, multi-wavelength sources combined with weight bands [46-52] or wavelength-sensitive elements [53-55] can be used to achieve synaptic connections within the wavelength domain. Wavelength division multiplexing (WDM) techniques are critical to fully exhaust the wideband advantages of optics. While the loaded data generally occupies an electronic bandwidth < 50 GHz limited by analog-to-digital converters and generic opto-electronic interfaces, the >30 THz optical bandwidth needs to be fully exploited by introducing multiple parallel wavelength channels to cover the full optical bandwidth.

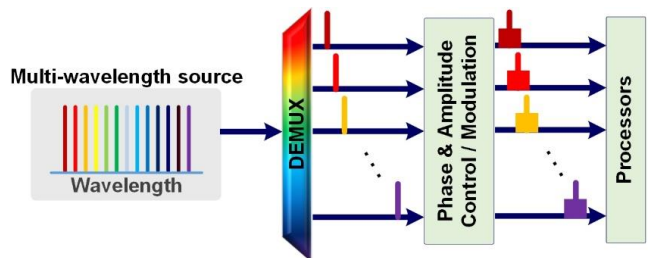

Figure 1. The overall architecture of a typical WDM system.

Optical frequency combs (OFCs), offering a large number of equally spaced wavelength channels, are powerful tools for communications systems and neuromorphic computing in order to significantly enhance the capacity or parallelism of the system [65-77]. Classic methods for OFC generation, such as those based on discrete laser arrays or electro-optic (EO) modulation, face limitations of one form or another, such as the bulky size and high cost, or the limited number of wavelengths. Kerr optical frequency combs [78-83], or "microcombs", originated from the parametric oscillation within high Q nonlinear micro-resonators, are promising on-chip OFC sources, as they can offer a large number of wavelengths in an integrated platform with much smaller footprint as well as higher scalability, performance, and reliability. To date, microcombs have enabled significant breakthroughs in frequency synthesis [128, 129], ultrahigh capacity communications [130-134], spectroscopy and measurement [135-144], microwave photonics [145-154], and neuromorphic optics [52, 54, 55].

In this paper, we review recent progress of the use of microcombs for neuromorphic optics. We introduce a convolution accelerator based on time-wavelength interleaving that operates at 11 Tera operations per second (TOPS), and discuss the potential and challenges for ONNs. The paper is structured as follows. Section II presents recent advances in optical frequency combs and microcombs, in terms of device platforms, nonlinearity and oscillation states, and generation methods. Section III reviews recent work of ONNs based on temporal-, spatial- and wavelength-division multiplexing (SDM, TDM and WDM). Section IV presents the theory and operation principle of a WDM-based convolution accelerator and experimental results of human image processing. The open challenges and limitations of ONNs are discussed in Section IV.

## II. INTEGRATED OPTICAL FREQUENCY COMBS FOR WAVELENGTH-DIVISION MULTIPLEXING

As shown in Figure 1, typical WDM systems consist of a multi-wavelength source that establishes parallel wavelength channels carrying either identical or different data for subsequent data transmission or processing, followed by wavelength multiplexers and demultiplexers that enable different wavelength channels to be modulated or processed separately.

OFCs are key to implementing the multi-wavelength sources in these systems, due to their relatively compact architecture with their inherent capability to offer equal frequency intervals between the comb lines that enables the easy frequency domain manipulation of the wavelength channels, [65-77] in contrast to discrete laser arrays. During the past two decades, the advance of photonic nanofabrication techniques has led to the production of integrated OFCs in different forms, offering remarkable advantages in terms of the system's size, weight, power consumption, and cost [77]. Existing integrated OFCs can be divided into several categories (Figures 2, 3) based on the underlying physical origins, including: a) Kerr frequency combs, or microcombs [78-83], that originate from parametric oscillation in an integrated micro-ring resonator (MRR); b) mode-locked lasers that employ gain media, such as Erbium-doped fiber amplifiers, in order to sustain oscillations and mode-locking mechanisms, such as saturable absorbers to yield pulsed outputs [84]; and c) electro-optically generated microcombs that employ modulators to take advantage of their second-order nonlinearity to introduce sidebands centered around a certain optical carrier [85].

For neuromorphic optics, the three categories of OFCs each have their own unique advantages and disadvantages. EO combs are more flexible in terms of tuning the comb spacing, although with the additional costs of external RF sources and oscillators. Mode-locked lasers support turn-key operation,

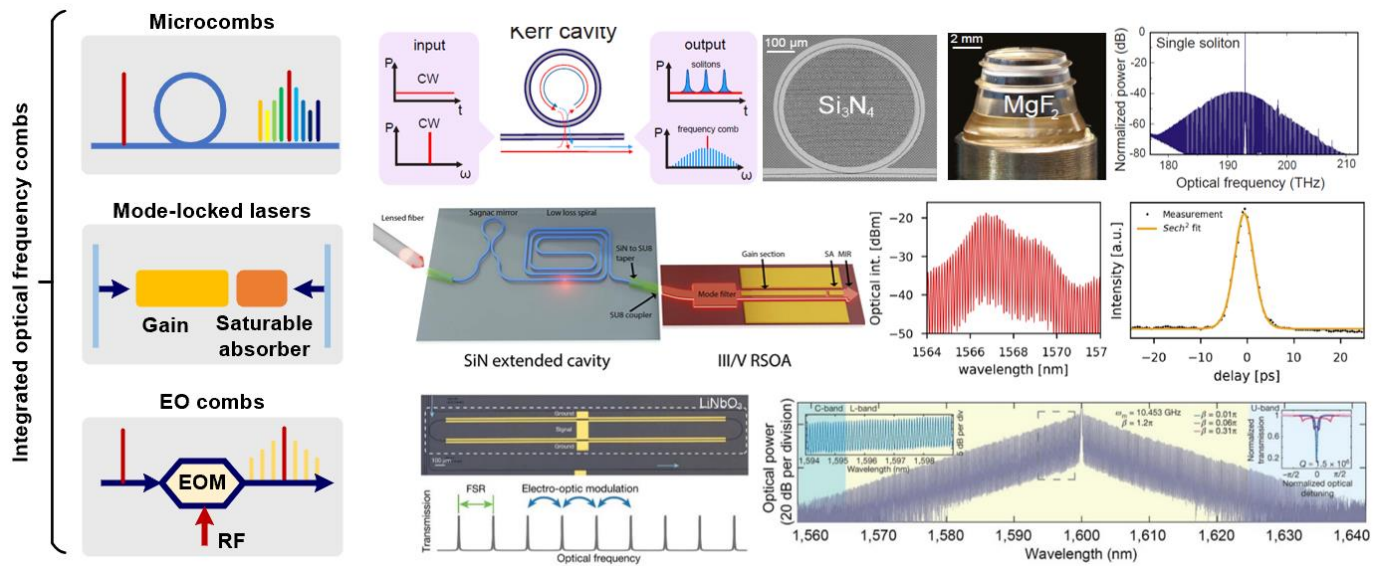

Figure 2. Integrated optical frequency combs. Figures are adapted from [84, 85, 103]

## Material platforms

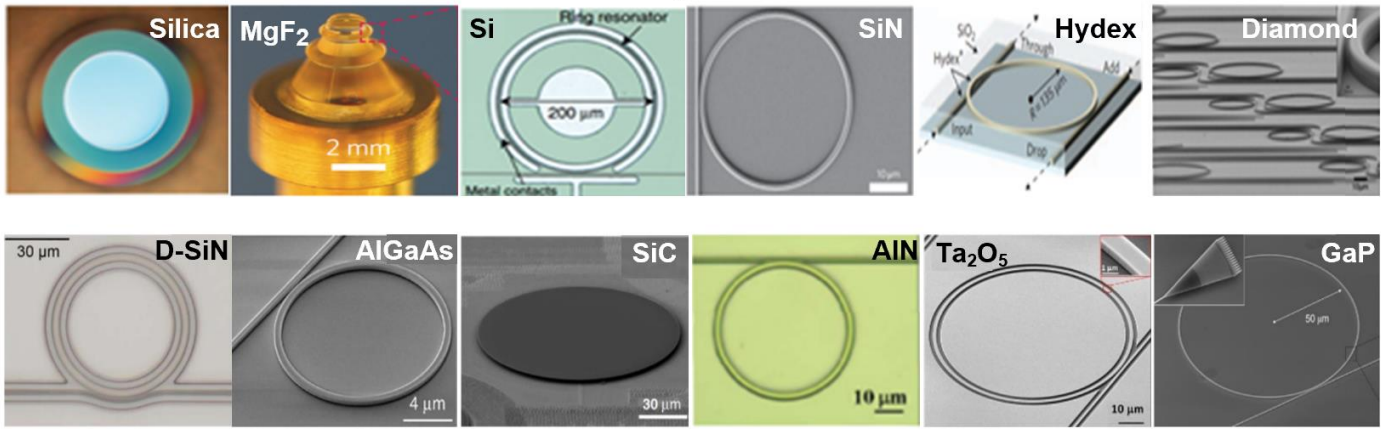

Figure 3. Material platforms of microcombs, including silica [88], MgF2 [89], silicon [90], silicon nitride [87], Hydex [92], diamond [93], deuterated silicon nitride [94], aluminum gallium arsenide [95], silicon carbide [97], aluminium nitride [98], tantalum pentoxide [99] and gallium phosphide [100]. Figures are adapted from [88-100].

offering low-noise coherent comb lines; however, their bandwidths are limited by the gain bandwidth of optical amplifiers (eg., Erbium-doped fiber amplifiers that operate in the C band from 1535-70 nm), limiting the computing parallelism. Microcombs, on the other hand, arguably offer the greatest advantages to enhance the performance of ONNs.

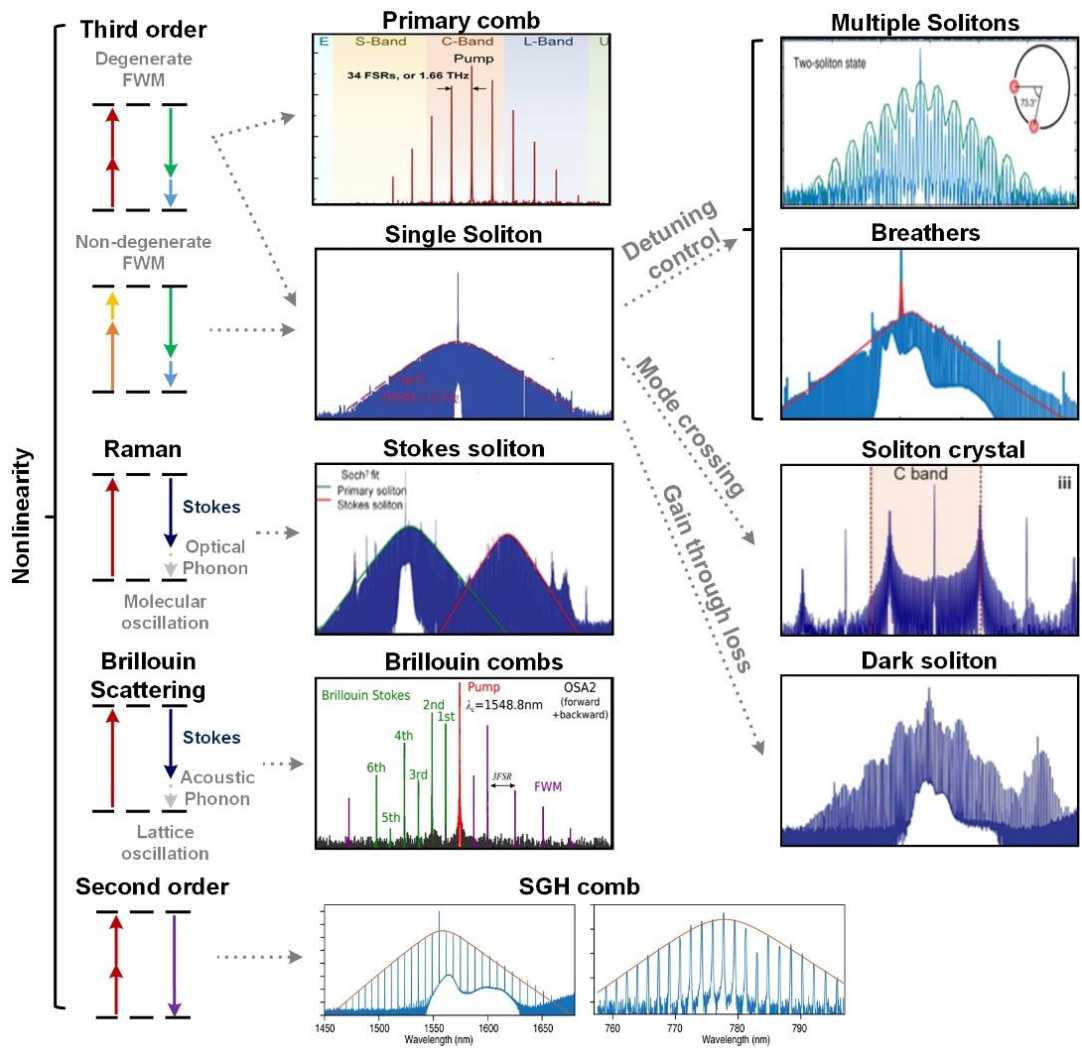

Figure 4. Diverse states of microcombs based on different types of nonlinear effects. Figures are adapted from [89, 103, 105, 113-115, 117, 146].

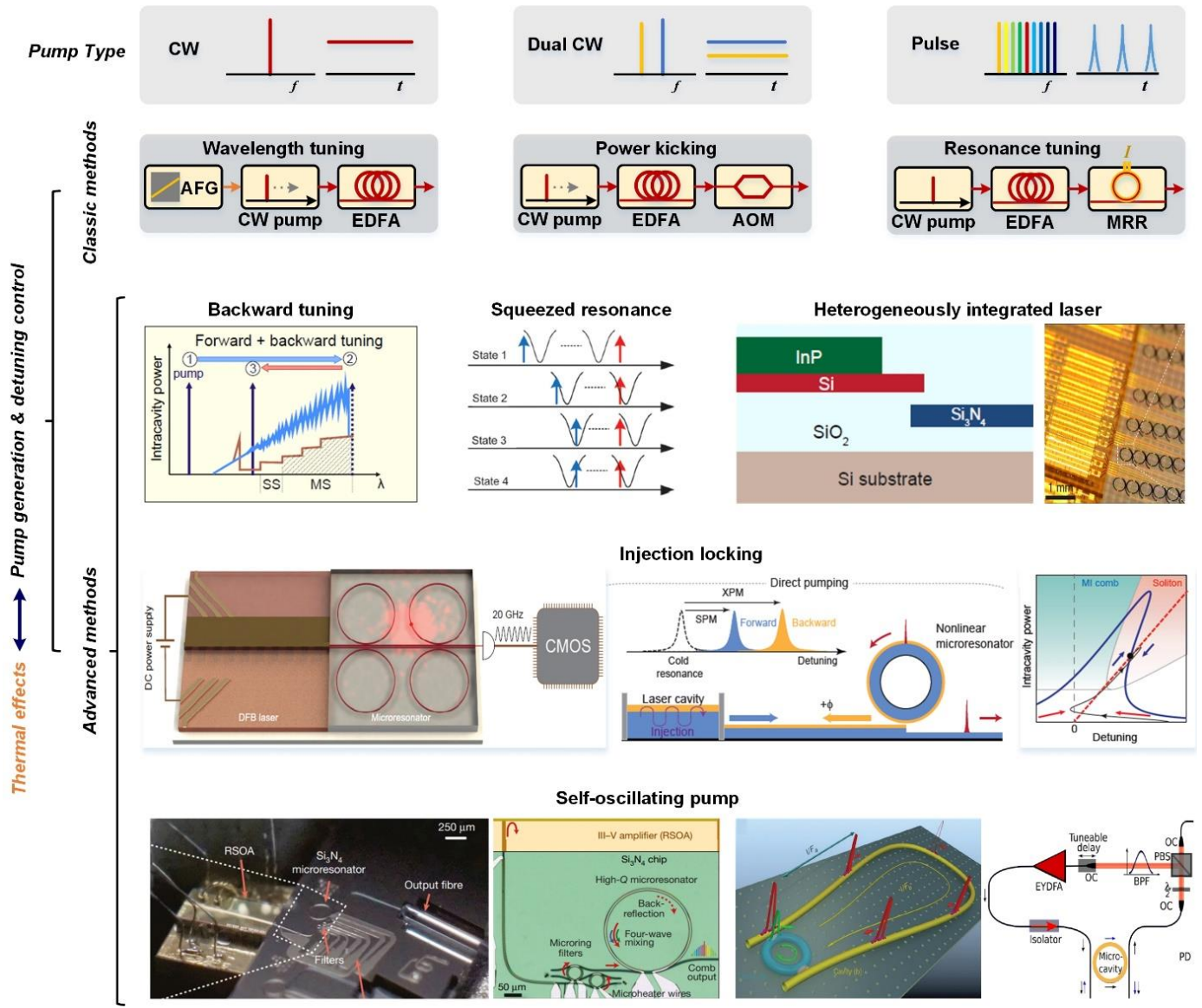

Figure 5. Pumping methods of microcombs. Figures are adapted from [103, 120, 124-127].

Microcombs are powerful integrated OFC sources, due to their compact footprint and ultra-wide bandwidths capable of octave-spanning operation, supported by broadband nonlinear parametric gain [78-83]. Microcombs originate from parametric oscillation within high-Q micro-resonators, which can be realized either in integrated form, such as micro-ring resonators [83], or in 3-dimensional form, such as spheres or rods [78]. The key to microcomb generation is to obtain sufficiently high parametric gain, which is directly determined by the strength of the third-order nonlinearity of the material platform and the Q factor of the resonator (i.e., low linear and nonlinear losses) [79]. For high Q-factor micro-resonators, the optical intra-cavity field can be significantly resonantly enhanced in order to initialize nonlinear phenomena that otherwise would generally require high optical power — such as modulation instability gain and parametric oscillation. In 2010, parametric oscillation based on integrated platforms was first reported [83, 86, 87], which revealed the ultimate potential of microcombs to be mass produced together with other optical components on a single chip, employing well-established CMOS platform fabrication techniques.

While microcomb generation often requires a high external pump power that brings about limitations in terms of energy efficiency and footprint (i.e., high-power amplifiers are unavoidably needed), significant effort has been made to reduce the parametric oscillation threshold. On the one hand, novel material platforms [88-100], such as SiC [97] and AlGaAs [95, 96] can exhibit significantly higher third-order optical nonlinearities, while on the other hand, advances in nanofabrication techniques, such as the Damascene reflow process and multi-mode waveguides [101-102], increase the Q factors to enhance the build-up optical fields in the micro-resonators. To date, a nonlinear coefficient of $n_2 = 2.6 \times 10^{-17}$ $m^2 W^{-1}$ has been achieved with AlGaAs waveguides, resulting in an ultra-low threshold power of 0.036 mW [95]. Q factors of over 10 million have been achieved with the Damascence reflow process for integrated micro-resonators [101], and these advances indicate that microcombs can be directly generated using a generic pump laser source where high-power amplifiers are no longer needed, albeit specific pump detuning control mechanisms are still necessary.

Governed by the Lugiato-Lefever equation [79], multiple parameters contribute to the rich dynamics of microcombs (Figure 4). These include the pump power and detuning (perturbed by thermal effects) that determine the intra-cavity pump power as well as the nonlinearity that governs the oscillation threshold and comb states. Also important is the dispersion that balances the nonlinearity, which together affect the comb bandwidth. Different nonlinear effects and their impact on microcomb generation have been investigated, including the third-order nonlinearity (degenerate and non-degenerate four-wave mixing, or FWM) that lead to primary combs, also termed Turing patterns (with comb spacings at multiple free spectral ranges (FSRs) of the micro-resonator). Using delicate pump detuning control, single solitons [89], multiple solitons [103, 104], and breathers [105-108] can be generated. On the other hand, mode crossings, can lead to soliton crystals [109-112] which display a range of attractive features, such as ease of generation and more stable and efficient operation. Finally, using a combination of gain and loss, normal dispersion and mode crossings can lead to dark solitons that offer similar advantages to soliton crystals, such as higher energy output [113]. Combs based on Raman scattering, a molecular-scale process where a Stokes photon and an optical photon are generated from a pump photon, can enable broadband Raman gain and Stokes solitons coexisting with Kerr combs [114]. Brillouin scattering, a lattice-scale process where a backward scattering Stokes photon and an acoustic phonon are generated from a pump photon, that introduces narrowband Brillouin gain at ~10 GHz away from the pump. As such, Brillouin combs are observed in large resonators with FSRs matching the Brillouin gain bandwidth[115, 116]. Finally, making use of the second-order nonlinearity can lead to the generation of frequency-doubled combs, with tailored dispersion to achieve phase matching at both oscillating wavelengths [117].

In parallel with the development of microcomb material platforms, advanced pumping methods have been demonstrated aiming at, on the one hand, overcoming the thermal effects of the micro-resonator that severely perturb the wavelength detuning control for soliton generation, and on the other hand, further reducing the complexity and footprint of the overall comb setup. Typically, the pump laser can take a number of different forms, including continuous-wave (CW) lasers that operate at a single wavelength, [89, 103, 118, 119] dual CW lasers with one serving as an auxiliary laser [120, 121]; and with optical pulses that feature much higher peak power to initialize parametric oscillation [122, 123].

Soliton generation requires that the pump wavelength be swept from the blue to red shifted side of the microresonator's resonance, finally landing in the soliton step region, during which the thermal effects of the micro-ring resonator shift the resonance with respect to the intra-cavity power [89] (Figure 5). Since the single soliton state features a much lower intra-cavity power compared to the originating chaotic state, deterministic soliton generation remains challenging and requires delicate external control of the pump-resonance detuning because of the inherent resonance shift that happens at the onset of soliton generation. Classic detuning control methods that have been widely employed and verified include fast wavelength/resonance tuning to reduce the accumulated heat during the detuning sweeping process [103, 119], as well as power kicking methods to manipulate the resonance shift induced by thermal effects and the optical pump power [108]. Recently, advances have been made, following the development of hybrid integration techniques, including more delicate control of the detuning in both forward and backward directions that lead to accurate control of the single and multiple soliton states [89]. Dual pump approaches that employ an auxiliary laser to balance the thermal effects induced by the pump laser [120, 121] have been successful. Heterogeneously integrated laser and micro-resonators will eventually enable mass production of the microcomb system [124], while injection locking approaches that lock the detuning via cross- and self-phase modulation effects to achieve turnkey microcomb generation [125-126] have been very successful. Finally, self-oscillating pump generation methods significantly reduce the pump power and enable high energy efficiency laser cavity soliton states [127].

All of these significant advances have collectively led to microcombs that exhibit an ever-increasing maturity for practical applications, providing a wideband, high-energy-efficiency, compact, turnkey and mass-producible comb source for WDM systems [128-157].

## III.   OPTICAL NEURAL NETWORKS BASED ON WAVELENGTH-DIVISION MULTIPLEXING

While electronic neuromorphic hardware faces increasingly large gaps between the desired and achievable performance in terms of computing density and energy efficiency, bounded by Moore's law, optical neuromorphic accelerators have attracted significant interest over the past decade, mainly due to their ultrawide optical bandwidth and low power consumption enabled by their inherently analog architectures [17-25].

The key to accelerate computing for artificial intelligence applications is to achieve the basic mathematical operations optically in a highly controlled manner, such that the overall parallelism and data throughput can be significantly enhanced for high computing power and large-scale fan-in/-out. High data throughput can be achieved with high-speed electro-optical interfaces, including electro-optical modulators and photodetectors that can reach over 50 GHz in bandwidth. High parallelism, needed to process large-scale data such as images or speech, can be implemented with multiplexing techniques that have been widely used in optical communications [17-18]. The parallelism of optical neural networks mainly determines the fan-in/out of the network, which governs the networks' capability of processing large-scale data. Likewise, the number of connections between the neurons denotes the networks' capability in terms of processing complicated tasks. Typical techniques to enhance the parallelism employ space-, time-, or wavelength-division multiplexing (Figure 6). ONNs based on space-division multiplexing (SDM) have been achieved with integrated photonic circuits [36-39], diffractive lens and others [40-45], where parallel input nodes are realized spatially in the form of different optical waveguide ports or pixels of the lens. Matrix multiplication operations have been achieved using intensity modulation/loss management of the optical field and mutual interference paths. ONNs based on time-division multiplexing (TDM) convert the input data and/or synaptic weights into temporal waveforms for matrix/vector operations [26-35], which requires additional optical buffers to achieve the accumulation process. The advantage of TDM techniques is that they are potentially capable of updating the synaptic weight at high speed in order to significantly accelerate the training process for the neural networks. However, they generally require other multiplexing techniques to enhance the computing power, due to their inherent sequential/serial operational methods.

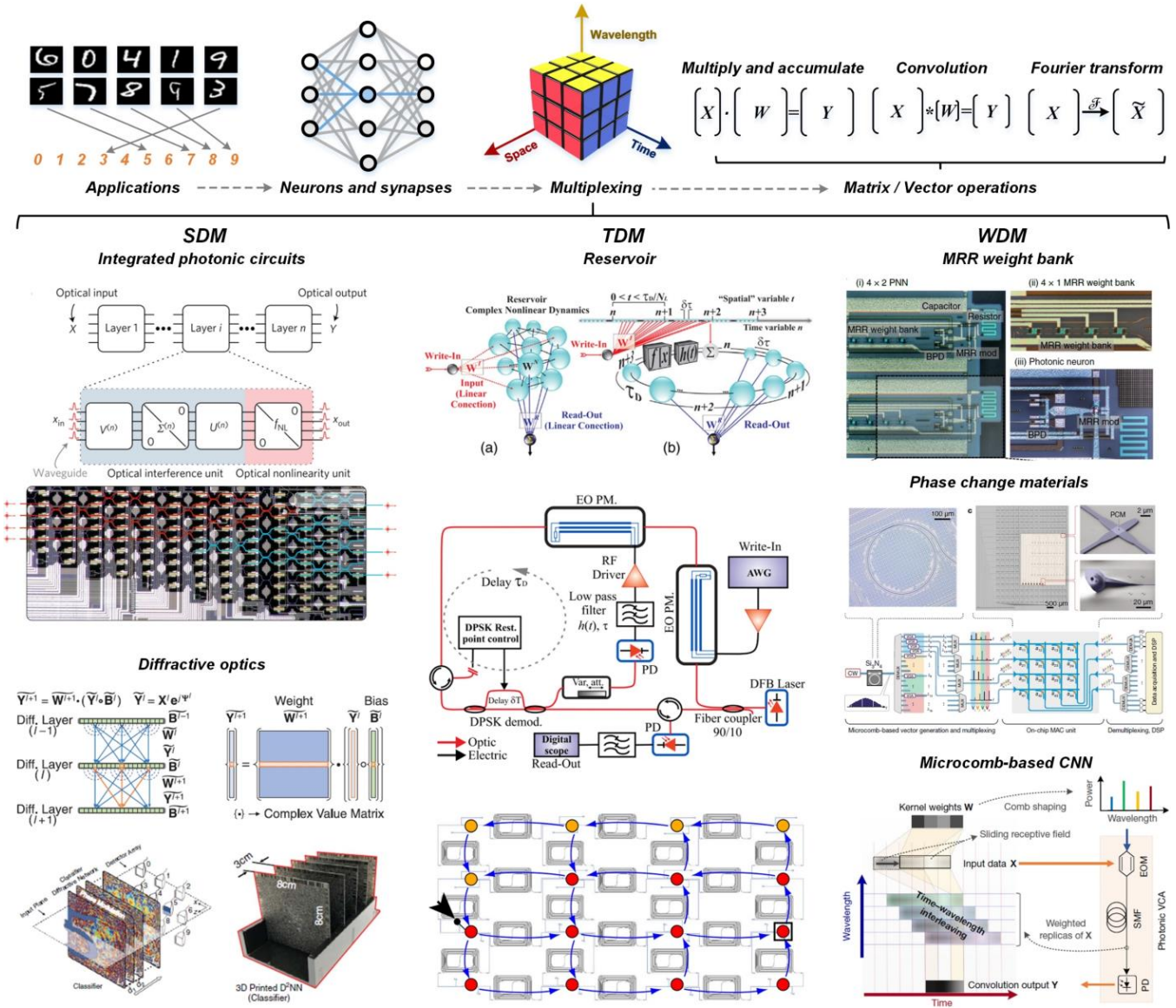

Figure 6. ONNs based on different multiplexing techniques. Figures are adapted from [28, 31, 36, 37, 40, 46, 52, 54]

Wavelength-division multiplexing (WDM) is a unique technique enabled by optics, which offers many advantages over purely electronic methods. Supported by the ultra-wide optical bandwidths up to 10s' of THz, 100's of wavelength channels can be established for parallel data processing of neural networks, thus leading to significantly enhanced computing speed – similar to the significantly enhanced data transmission capacity for WDM-based communications systems. Current WDM-based ONNs [46-55] can be generally categorized into two types, according to whether the wavelength channels carry identical or different data. Individually modulated wavelength channels [46, 52] enable potentially higher flexibility and parallelism, capable of performing generic matrix multiplication operations. However, they also require large arrays of modulators with well-matched wavelength multiplexers/demultiplexers, which remains challenging to be integrated and synchronized on chip without significant increase in complexity and cost. For example, for a 50GHz-spacing microcomb source, over 80 wavelength channels can be established in the C band, which in turn requires over 80 modulators. In parallel, simultaneously modulated wavelength channels [54, 55] are potentially much more straightforward to implement and integrate, as only one modulator is needed to broadcast the input data. The drawback to this approach is that, despite the increase in parallelism, the overall computing power is limited for generic matrix operations. This is true for all broadcast-and-weight systems used for matrix multiplication [55], where the input data is implemented as serial temporal waveforms. Here, WDM does not in fact lead to an enhancement in the computing parallelism and so the computing speed is similar to the data rate [54].

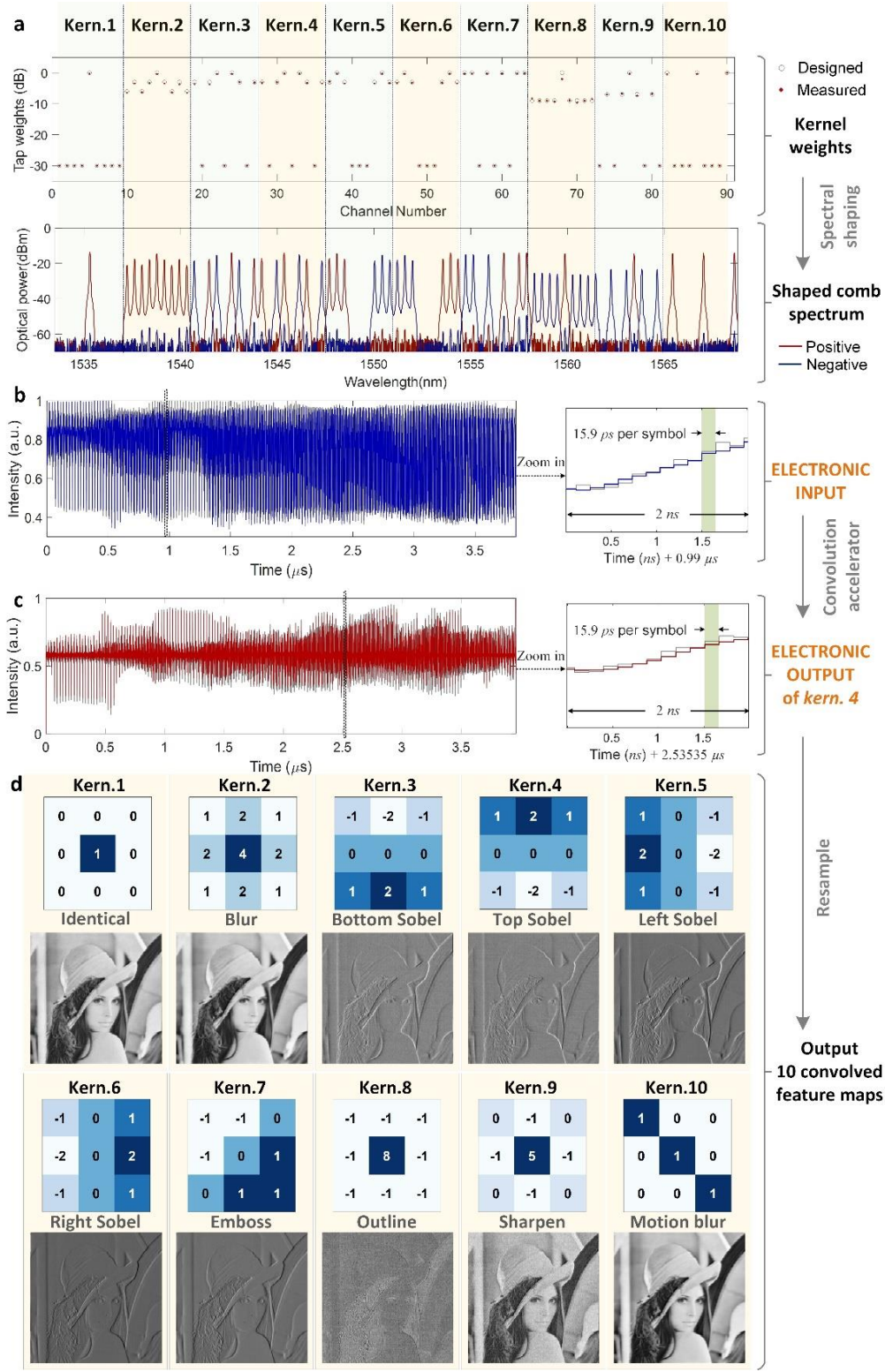

Figure 7. Experimental results of the convolution accelerator. a. The kernel weights and the shaped microcomb's optical spectrum. b. The input electrical waveform of the image (the grey and blue lines show the ideal and experimentally generated waveforms, respectively). c. The convolved results of the fourth kernel that performs top Sobel image processing function (the grey and red lines show the ideal and experimentally generated waveforms, respectively). d. The weight matrices of the employed kernels and corresponding recovered images.

## IV. WDM-BASED CONVOLUTION ACCELERATORS

Recently, [54] convolution accelerators have been proposed based on a time wavelength interleaving approach, which avoids the trade-offs mentioned above and has achieved high computing power within a compact footprint. The operation principle of the convolution accelerator is illustrated in Figure 7. For vector convolutions between a 1×L data vector and a 1×R weight vector, the data vector is converted to a temporal waveform $\mathbf{X}[n]$ via digital-to-analog converters, where n denotes discrete temporal locations of the symbols. The weight vector is then mapped onto the power of optical comb lines as $\mathbf{W}[R-i+1]$ ($1 \leqslant i \leqslant R$, i increases with wavelength). Via electro-optical modulation, $\mathbf{X}[n]$ can be broadcast onto all of the comb lines simultaneously, yielding weighted replicas as $\mathbf{W}[R-i+1] \cdot \mathbf{X}[n]$. Next, the weighted replicas are progressively delayed via dispersion, with the delay step between adjacent wavelength channels equaling to the symbol duration of $\mathbf{X}[n]$, thus yielding delayed replicas $\mathbf{W}[R-i+1] \cdot \mathbf{X}[n-i]$. After photodetection, the replicas are summed as

$$\mathbf{Y}[n] = \sum_{i=1}^{R} \mathbf{W}[R-i+1] \cdot \mathbf{X}[n-i] = (\mathbf{W} * \mathbf{X})[n] \quad (1)$$

where each symbol $\mathbf{Y}[n]$ within the range of [R+1, L+1] denotes the dot product between $\mathbf{W}$ and a sliding window [n−R, n−R+1, n−R+2, …, n−1] of $\mathbf{X}$, thus achieving convolution operations between the input and weight vectors.

The output waveform $\mathbf{Y}$ contains R+L-1 symbols, amongst which R-L+1 symbols denote the convolution results, for each computing cycle. Here, each symbol is the result of R multiplications and R accumulation operations. As such, the computing speed can be given as (R+L-1)/(R-L+1)×2×R× B, where B denotes the symbol rate. For practical applications, the length of the input data vector is much larger than that of the weight vector (L >> R), thus (R+L-1)/(R-L+1) ≈ 1 and the computing speed can be given as 2×R×B.

In addition, the same hardware architecture can be tailored to achieve matrix multiplication for the fully connected layer. Assuming that the input data vector $\mathbf{X_{FC}}[n]$ and the weight vector $\mathbf{W_{FC}}[R-i+1]$ both have a length of R ($1 \leqslant i \leqslant R$, $1 \leqslant n \leqslant R$), thus, according to Equation 1, the output waveform after photodetection is

$$\mathbf{Y_{FC}}[n] = \sum_{i=1}^{R} \mathbf{W_{FC}}[R-i+1] \cdot \mathbf{X_{FC}}[n-i] \quad (2)$$

By sampling at the time slot denoted by n=R+1, the matrix multiplication result of the two input vectors is obtained as

$$\begin{aligned} \mathbf{Y_{FC}}[R+1] &= \sum_{i=1}^{R} \mathbf{W_{FC}}[R-i+1] \cdot \mathbf{X_{FC}}[R+1-i] \\ &= \sum_{i=1}^{R} \mathbf{W_{FC}}[i] \cdot \mathbf{X_{FC}}[i] \end{aligned} \quad (3)$$

The output waveform $\mathbf{Y}$ contains 2L-1 symbols for each computing cycle, amongst which only one symbol denotes the vector multiplication result from L multiplications and L accumulation operations. As such, the computing speed can be given as 1/(2L+1)×2×L×B ≈ B.

As illustrated above, this time wavelength interleaving method can significantly enhance the computing power with WDM and a single modulator, albeit it is applicable only for specific convolution operations, rather than general matrix operations. Figure 7 shows the experimental results for an 11 Tera-Ops convolutional accelerator processing large scale 500x500 pixel facial images.

The convolution accelerator can also be used to form convolutional neural networks (Figure 8), which has shown unprecedented performance for image recognition applications. Here, we present the results of human image processing, using the convolutional accelerator as illustrated in [54]. Here, 90 comb lines were employed to form ten 3×3 convolutional kernels, achieving diverse image processing functions, including: identical, blur, bottom/top/left/right Sobel, emboss, outline, sharpen and motion blur. A 500 × 500 input image was flattened into a vector and converted into an electrical input waveform via a high-speed electrical digital-to-analog converter, at a data rate of 62.9 Giga Baud. The waveform was then broadcast onto all wavelength channels and weighted via electro-optical modulation. Following this, the weighted replicas were transmitted through ~2.2 km of standard single mode fibre (dispersion ~17ps/nm/km), which corresponds to a progressive delay of 15.9 ps that matches with the symbol rate. The wavelength channels were then de-multiplexed into 10 sub-bands that corresponded to the 10 convolutional kernels, and finally summed upon photodetection. The output waveforms were finally sampled and rescaled to form the feature maps, denoting a diverse range of pre-defined hierarchical features of the input image. In combination with a fully connected layer, the convolutional accelerators can be used to form convolutional neural networks, where the kernels' weights are trained for specific tasks/datasets such as facial recognition.

## V. DISCUSSIONS

While ONNs show great potential in achieving high computing power and energy efficiency, their inherent analog framework indicates that ultimately, hybrid opto-electronic neuromorphic hardware is a likely optimal solution that takes advantage of both the broadband optics and the versatility of digital electronics, where optics undertakes the majority of the computing operations and electronics manages the date flow and storage. In such architectures, particularly for deep learning networks with many internal hidden layers, optical computing units will be iteratively introduced, linked by electronics, to perform certain computing functions for each network layer, while electronics will likely still be needed to manage the overall network logistics, including the structure of the network, linking the hidden layers, and the data/parameter flow. Having said this, though, we note that there is still significant room to increase the range of functions performed optically, that currently tend to be done using

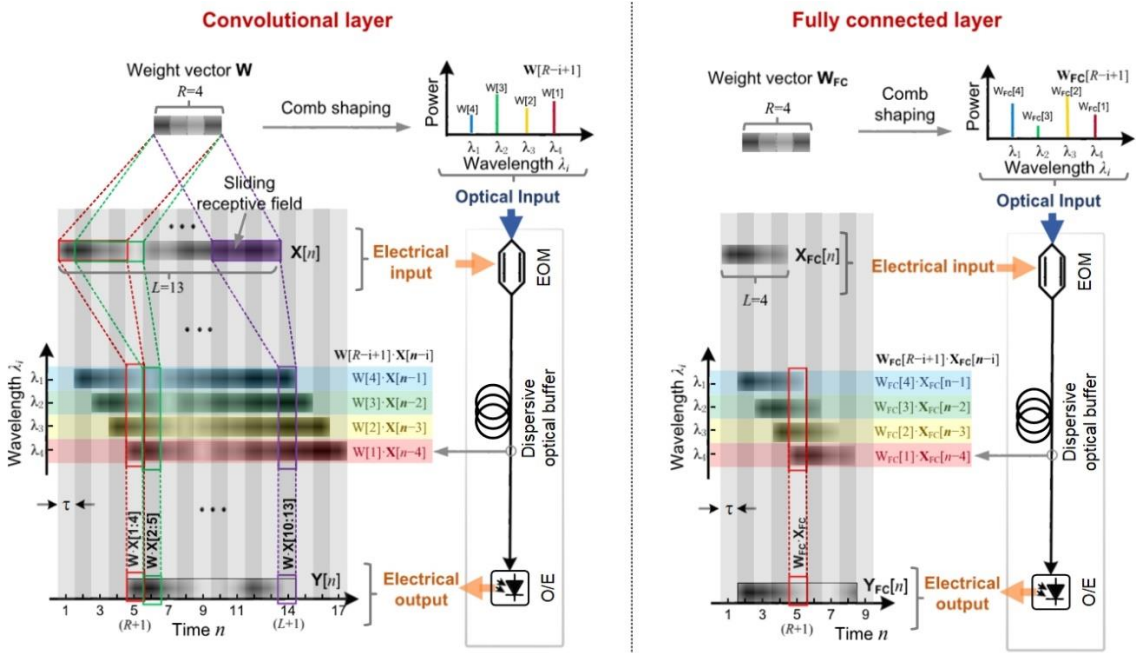

Figure 8. The operation principle of the photonic convolution accelerator for the convolutional layer with R = 4 and L = 13, and the fully connected layer with R = L = 4, consisting of an electro-optical modulator (EOM), an optical buffer that has progressive wavelength-sensitive delay, and an optical-to-electrical conversion module (O/E). Figures are adapted from [54].

electronics. These include the pooling layer as well as the output nonlinearity.

So far, although notable progress has been made on ONNs, many challenges still exist that need to be addressed for future applications. First, dense integration of the entire photonics system needs to be achieved, as this is the key to achieve competitive computing parallelism for ONNs in comparison to their electrical counterparts. Ultimately achieving computing with millions of parameters — sufficient for all practical applications — remains challenging for ONNs based on SDM or WDM, since the parameters need to be mapped onto physical divisions. TDM techniques can potentially address these needs, as the parameters are mapped onto temporal waveforms that theoretically can have infinite lengths, such as convolving vectors with a length of 0.25 million using TDM. [54] For WDM methods, architectures have been proposed [54] that are able to scale up to 25,000 synapses using standard off-the-shelf telecommunications equipment.

In parallel, hybrid integration techniques, capable of integrating the comb source and subsequent components, including in particular spectral shapers (utilizing high-resolution wavelength demultiplexers and arrays of amplitude/phase controlling units achieved via either passive or active components), are necessary to make optimal use of optics' broad bandwidths with WDM techniques

Secondly, more categories of computing operations (such as the nonlinear functions of neurons and Fourier transforms) and network architectures (such as graph neural networks [158]) need to be demonstrated on chip to further enhance the universality of ONNs for diverse machine learning tasks. This will rely on advances in both novel computing architectures tailored to specific operations and the integration of high-nonlinearity components that can realize the nonlinear functions with relatively low optical power.

Thirdly, as ONNs will be achieved as assembles of massive programmable photonic units for a high spatial-division parallelism, tailored algorithms to overcome the challenges of fabrication imperfections and on-chip cross-talk are necessary for fast-converging control of on-chip elements and training of the networks.

Finally, the unique advantages of optics, especially in terms of its compatibility with general optical devices such as cameras [159] or gratings [160] for image processing, should be investigated to further boost ONNs' performance for these applications. Since these architectures avoid the need of digital electronics at the frontend (for data format conversion such as electrical-to-optical and analog-to-digital), the power consumption can be greatly reduced.

With these challenges being fully addressed, ONNs can then be plugged into existing electronic hardware to significantly enhance the computing performance of the whole system, dramatically accelerating the training speed of computationally intense neural networks, and thus in turn potentially lead to more complicated and intelligent networks for advanced machine learning tasks such as fully automated vehicles and real-time image/video processing.

## VI. CONCLUSION

We have reviewed recent advances in WDM-based ONNs, focusing on methods that use integrated microcombs to implement ONNs. We present results for human image processing using an optical convolution accelerator operating at 11 Tera operations per second. The open challenges and limitations of ONNs that need to be addressed for future applications are also discussed.

## REFERENCES

1. Y. LeCun, Y. Bengio, and G. Hinton, "Deep learning," Nature, vol. 521, no. 7553, pp. 436-444, May 28, 2015.
2. A. Krizhevsky, I. Sutskever, and G. E. Hinton, "ImageNet Classification with Deep Convolutional Neural Networks," Communications of the Acm, vol. 60, no. 6, pp. 84-90, Jun, 2017.
3. D. Silver et al., "Mastering the game of Go with deep neural networks and tree search," Nature, vol. 529, no. 7587, pp. 484-+, Jan 28, 2016.
4. M. Reichstein et al., "Deep learning and process understanding for data-driven Earth system science," Nature, vol. 566, no. 7743, pp. 195-204, Feb 14, 2019.
5. P. M. R. DeVries, F. Viegas, M. Wattenberg, and B. J. Meade, "Deep learning of aftershock patterns following large earthquakes," Nature, vol. 560, no. 7720, pp. 632-+, Aug 30, 2018.
6. M. Reichstein et al., "Deep learning and process understanding for data-driven Earth system science," Nature, vol. 566, no. 7743, pp. 195-204, Feb 14, 2019.
7. S. Webb, "Deep Learning for Biology," Nature, vol. 554, no. 7693, pp. 555-557, Feb 22, 2018.
8. A. Esteva et al., "Dermatologist-level classification of skin cancer with deep neural networks (vol 542, pg 115, 2017)," Nature, vol. 546, no. 7660, pp. 686-686, Jun 29, 2017.
9. D. Capper et al., "DNA methylation-based classification of central nervous system tumours," Nature, vol. 555, no. 7697, pp. 469-+, Mar 22, 2018.
10. V. Mnih et al., "Human-level control through deep reinforcement learning," Nature, vol. 518, no. 7540, pp. 529-533, Feb 26, 2015.
11. X. W. Xu et al., "Scaling for edge inference of deep neural networks," Nature Electronics, vol. 1, no. 4, pp. 216-222, Apr, 2018.
12. H. Sutter, "The free lunch is over: A fundamental turn toward concurrency in software," Dr Dobb's J., 30, 202–210 (2005).
13. C. Toumey, "Less is Moore," Nature Nanotechnology, vol. 11, no. 1, pp. 2-3, Jan, 2016.
14. R. G. Dreslinski, M. Wieckowski, D. Blaauw, D. Sylvester, and T. Mudge, "Near-Threshold Computing: Reclaiming Moore's Law Through Energy Efficient Integrated Circuits," Proceedings of the Ieee, vol. 98, no. 2, pp. 253-266, Feb, 2010.
15. S. Ambrogio et al., "Equivalent-accuracy accelerated neural-network training using analogue memory," Nature, vol. 558, no. 7708, pp. 60-+, Jun 7, 2018.
16. D. A. B. Miller, "Attojoule Optoelectronics for Low-Energy Information Processing and Communications," Journal of Lightwave Technology, vol. 35, no. 3, pp. 346-396, Feb 1, 2017.
17. B. J. Shastri et al., "Photonics for artificial intelligence and neuromorphic computing," Nature Photonics, vol. 15, no. 2, pp. 102-114, Feb, 2021.
18. G. Wetzstein et al., "Inference in artificial intelligence with deep optics and photonics," Nature, vol. 588, no. 7836, pp. 39-47, Dec 3, 2021.
19. C. Huang et al., "Prospects and applications of photonic neural networks," Advances in Physics-X, vol. 7, no. 1, Jan 1, 2022.
20. H. L. Zhou et al., "Photonic matrix multiplication lights up photonic accelerator and beyond," Light-Science & Applications, vol. 11, no. 1, Feb 3, 2022.
21. K. Berggren et al., "Roadmap on emerging hardware and technology for machine learning," Nanotechnology, vol. 32, no. 1, pp. 012002, Jan 1, 2021.
22. T. Ferreira de Lima et al., "Primer on silicon neuromorphic photonic processors: architecture and compiler," Nanophotonics, vol. 9, no. 13, pp. 4055-4073, 2020.
23. E. Goi, Q. M. Zhang, X. Chen, H. T. Luan, and M. Gu, "Perspective on photonic memristive neuromorphic computing," Photonix, vol. 1, no. 1, Mar 3, 2020.
24. T. F. de Lima, B. J. Shastri, A. N. Tait, M. A. Nahmias, and P. R. Prucnal, "Progress in neuromorphic photonics," Nanophotonics, vol. 6, no. 3, pp. 577-599, May, 2017.
25. Q. M. Zhang, H. Y. Yu, M. Barbiero, B. K. Wang, and M. Gu, "Artificial neural networks enabled by nanophotonics," Light-Science & Applications, vol. 8, May 8, 2019.
26. L. Appeltant et al., "Information processing using a single dynamical node as complex system," Nat Commun, vol. 2, pp.

27. F. Duport, B. Schneider, A. Smerieri, M. Haelterman, and S. Massar, "All-optical reservoir computing," Optics Express, vol. 20, no. 20, pp. 22783-22795, Sep 24, 2012.
28. L. Larger et al., "High-Speed Photonic Reservoir Computing Using a Time-Delay-Based Architecture: Million Words per Second Classification," Physical Review X, vol. 7, no. 1, Feb 6, 2017.
29. Y. Paquot et al., "Optoelectronic reservoir computing," Sci Rep, vol. 2, pp. 287, 2012.
30. D. Brunner, M. C. Soriano, C. R. Mirasso, and I. Fischer, "Parallel photonic information processing at gigabyte per second data rates using transient states," Nat Commun, vol. 4, pp. 1364, 2013.
31. K. Vandoorne et al., "Experimental demonstration of reservoir computing on a silicon photonics chip," Nat Commun, vol. 5, pp. 3541, Mar 24, 2014.
32. B. Romeira, R. Avo, J. M. Figueiredo, S. Barland, and J. Javaloyes, "Regenerative memory in time-delayed neuromorphic photonic resonators," Sci Rep, vol. 6, pp. 19510, Jan 19, 2016.
33. J. Bueno et al., "Reinforcement learning in a large-scale photonic recurrent neural network," Optica, vol. 5, no. 6, 2018.
34. J. Robertson, M. Hejda, J. Bueno, and A. Hurtado, "Ultrafast optical integration and pattern classification for neuromorphic photonics based on spiking VCSEL neurons," Sci Rep, vol. 10, no. 1, pp. 6098, Apr 8, 2020.
35. K. S. Kravtsov, M. P. Fok, P. R. Prucnal, and D. Rosenbluth, "Ultrafast all-optical implementation of a leaky integrate-and-fire neuron," Opt Express, vol. 19, no. 3, pp. 2133-47, Jan 31, 2011.
36. Y. Shen et al., "Deep learning with coherent nanophotonic circuits," Nature Photonics, vol. 11, no. 7, pp. 441-446, 2017.
37. J. Feldmann, N. Youngblood, C. D. Wright, H. Bhaskaran, and W. H. P. Pernice, "All-optical spiking neurosynaptic networks with self-learning capabilities," Nature, vol. 569, no. 7755, pp. 208-214, 2019.
38. H. Zhang et al., "An optical neural chip for implementing complex-valued neural network," Nature Communications, vol. 12, no. 1, Jan 19, 2021.
39. H. H. Zhu et al., "Space-efficient optical computing with an integrated chip diffractive neural network," Nature Communications, vol. 13, no. 1, Feb 24, 2022.
40. J. Chang, V. Sitzmann, X. Dun, W. Heidrich, and G. Wetzstein, "Hybrid optical-electronic convolutional neural networks with optimized diffractive optics for image classification," Sci Rep, vol. 8, no. 1, pp. 12324, Aug 17, 2018.
41. X. Lin et al., "All-optical machine learning using diffractive deep neural networks," Science, vol. 361, no. 6406, pp. 1004-+, Sep 7, 2018.
42. C. Qian et al., "Performing optical logic operations by a diffractive neural network," Light Sci Appl, vol. 9, pp. 59, 2020.
43. T. Wang et al., "An optical neural network using less than 1 photon per multiplication," Nature Communications, vol. 13, no. 1, pp. 123, 2022/01/10, 2022.
44. T. Yan et al., "Fourier-space Diffractive Deep Neural Network," Physical Review Letters, vol. 123, no. 2, Jul 9, 2019.
45. T. K. Zhou et al., "Large-scale neuromorphic optoelectronic computing with a reconfigurable diffractive processing unit," Nature Photonics, vol. 15, no. 5, pp. 367-373, May, 2021.
46. C. R. Huang et al., "A silicon photonic-electronic neural network for fibre nonlinearity compensation," Nature Electronics, vol. 4, no. 11, pp. 837-844, Nov, 2021.
47. M. A. Nahmias, B. J. Shastri, A. N. Tait, and P. R. Prucnal, "A Leaky Integrate-and-Fire Laser Neuron for Ultrafast Cognitive Computing," Ieee Journal of Selected Topics in Quantum Electronics, vol. 19, no. 5, Sep-Oct, 2013.
48. B. J. Shastri et al., "Spike processing with a graphene excitable laser," Sci Rep, vol. 6, pp. 19126, Jan 12, 2016.
49. A. N. Tait et al., "Microring Weight Banks," Ieee Journal of Selected Topics in Quantum Electronics, vol. 22, no. 6, Nov-Dec, 2016.
50. M. A. Nahmias et al., "Photonic Multiply-Accumulate Operations for Neural Networks," Ieee Journal of Selected Topics in Quantum Electronics, vol. 26, no. 1, Jan-Feb, 2020.
51. A. N. Tait et al., "Neuromorphic photonic networks using silicon photonic weight banks," Sci Rep, vol. 7, no. 1, pp. 7430, Aug 7, 2017.

52. J. Feldmann et al., "Parallel convolutional processing using an integrated photonic tensor core," Nature, vol. 589, no. 7840, pp. 52-+, Jan 7, 2021.

53. A. N. Tait, J. Chang, B. J. Shastri, M. A. Nahmias, and P. R. Prucnal, "Demonstration of WDM weighted addition for principal component analysis," Optics Express, vol. 23, no. 10, pp. 12758-12765, May 18, 2015.

54. X. Xu et al., "11 TOPS photonic convolutional accelerator for optical neural networks," Nature, vol. 589, no. 7840, p44, 2021.

55. X. Xu et al., "Photonic Perceptron Based on a Kerr Microcomb for High-Speed, Scalable, Optical Neural Networks," Laser & Photonics Reviews, vol. 14, no. 10, 2000070, Oct, 2020.

56. T. W. Hughes, M. Minkov, Y. Shi, and S. Fan, "Training of photonic neural networks through in situ backpropagation and gradient measurement," Optica, vol. 5, no. 7, 2018.

57. T. W. Hughes, I. A. D. Williamson, M. Minkov, and S. H. Fan, "Wave physics as an analog recurrent neural network," Science Advances, vol. 5, no. 12, Dec, 2019.

58. R. Hamerly, L. Bernstein, A. Sludds, M. Soljacic, and D. Englund, "Large-Scale Optical Neural Networks Based on Photoelectric Multiplication," Physical Review X, vol. 9, no. 2, May 16, 2019.

59. G. Mourgias-Alexandris et al., "An all-optical neuron with sigmoid activation function," Opt Express, vol. 27, no. 7, pp. 9620-9630, Apr 1, 2019.

60. Y. Zuo et al., "All-optical neural network with nonlinear activation functions," Optica, vol. 6, no. 9, 2019.

61. I. A. D. Williamson et al., "Reprogrammable Electro-Optic Nonlinear Activation Functions for Optical Neural Networks," Ieee Journal of Selected Topics in Quantum Electronics, vol. 26, no. 1, Jan-Feb, 2020.

62. B. Shi, N. Calabretta, and R. Stabile, "Deep Neural Network Through an InP SOA-Based Photonic Integrated Cross-Connect," IEEE Journal of Selected Topics in Quantum Electronics, vol. 26, no. 1, pp. 1-11, 2020.

63. E. Goi et al., "Nanoprinted high-neuron-density optical linear perceptrons performing near-infrared inference on a CMOS chip," Light-Science & Applications, vol. 10, no. 1, Mar 3, 2021.

64. H. Q. Deng, and M. Khajavikhan, "Parity time symmetric optical neural networks," Optica, vol. 8, no. 10, pp. 1328-1333, Oct 20, 2021.

65. D. J. Jones, S. A. Diddams, J. K. Ranka, A. Stentz, R. S. Windeler, J. L. Hall, and S. T. Cundiff, "Carrier-envelope phase control of femtosecond mode-locked lasers and direct optical frequency synthesis," Science, vol. 288, no. 5466, pp. 635-639, Apr 28, 2000.

66. S. A. Diddams, D. J. Jones, J. Ye, S. T. Cundiff, J. L. Hall, J. K. Ranka, R. S. Windeler, R. Holzwarth, T. Udem, and T. W. Hansch, "Direct link between microwave and optical frequencies with a 300 THz femtosecond laser comb," Physical Review Letters, vol. 84, no. 22, pp. 5102-5105, May 29, 2000.

67. R. Holzwarth, T. Udem, T. W. Hansch, J. C. Knight, W. J. Wadsworth, and P. S. J. Russell, "Optical frequency synthesizer for precision spectroscopy," Physical Review Letters, vol. 85, no. 11, pp. 2264-2267, Sep 11, 2000.

68. T. Udem, R. Holzwarth, and T. W. Hansch, "Optical frequency metrology," Nature, vol. 416, no. 6877, pp. 233-237, Mar 14, 2002.

69. B. Bernhardt, A. Ozawa, P. Jacquet, M. Jacquey, Y. Kobayashi, T. Udem, R. Holzwarth, G. Guelachvili, T. W. Hansch, and N. Picque, "Cavity-enhanced dual-comb spectroscopy," Nature Photonics, vol. 4, no. 1, pp. 55-57, Jan 10, 2010.

70. T. Ideguchi, A. Poisson, G. Guelachvili, N. Picque, and T. W. Hansch, "Adaptive real-time dual-comb spectroscopy," Nature Communications, vol. 5, Feb, 2014.

71. D. Hillerkuss et al., "26 Tbit s(-1) line-rate super-channel transmission utilizing all-optical fast Fourier transform processing," Nature Photonics, vol. 5, no. 6, pp. 364-371, Jun, 2011.

72. S. T. Cundiff, and A. M. Weiner, "Optical arbitrary waveform generation," Nature Photonics, vol. 4, no. 11, pp. 760-766, Nov, 2010.

73. T. M. Fortier, M. S. Kirchner, F. Quinlan, J. Taylor, J. C. Bergquist, T. Rosenband, N. Lemke, A. Ludlow, Y. Jiang, C. W. Oates, and S. A. Diddams, "Generation of ultrastable microwaves via optical frequency division," Nature Photonics, vol. 5, no. 7, pp. 425-429, Jul 2011.

74. V. R. Supradeepa, C. M. Long, R. Wu, F. Ferdous, E. Hamidi, D. E. Leaird, and A. M. Weiner, "Comb-based radiofrequency photonic filters with rapid tunability and high selectivity," Nature Photonics, vol. 6, no. 3, pp. 186-194, Mar, 2012.

75. V. Torres-Company, and A. M. Weiner, "Optical frequency comb technology for ultra-broadband radio-frequency photonics," Laser & Photonics Reviews, vol. 8, no. 3, pp. 368-393, May, 2014.

76. C. H. Chen, C He, D. Zhu, R. H. Guo, F. Z. Zhang, and S. L. Pan, "Generation of a flat optical frequency comb based on a cascaded polarization modulator and phase modulator," Optics Letters, vol. 38, no. 16, pp. 3137-3139, Aug 15, 2013.

77. L. Chang, S. T. Liu, and J. E. Bowers, "Integrated optical frequency comb technologies," Nature Photonics, vol. 16, no. 2, pp. 95-108, Feb, 2022.

78. P. Del'Haye, A. Schliesser, O. Arcizet, T. Wilken, R. Holzwarth, and T. J. Kippenberg, "Optical frequency comb generation from a monolithic microresonator," Nature 450, 1214-1217 (2007).

79. A. Pasquazi, M. Peccianti, L. Razzari, D. J. Moss, S. Coen, M. Erkintalo, Y. K. Chembo, T. Hansson, S. Wabnitz, and P. Del'Haye, "Micro-combs: A novel generation of optical sources," Physics Reports vol. 729, 1-81 (2018).

80. T. J. Kippenberg, A. L. Gaeta, M. Lipson, and M. L. Gorodetsky, "Dissipative Kerr solitons in optical microresonators," Science 361 (2018).

81. A. L. Gaeta, M. Lipson, and T. J. Kippenberg, "Photonic-chip-based frequency combs," Nat. Photonics 13, 158-169 (2019).

82. M. Kues, C. Reimer, J. M. Lukens, W. J. Munro, A. M. Weiner, D. J. Moss, and R. Morandotti, "Quantum optical microcombs," Nat. Photonics vol. 13, 170-179 (2019).

83. D. J. Moss, R. Morandotti, A. L. Gaeta, and M. Lipson, "New CMOS-compatible platforms based on silicon nitride and Hydex for nonlinear optics," Nat. Photonics 7, 597-607 (2013).

84. E. Vissers, S. Poelman, C. O. de Beeck, K. Van Gasse, and B. Kuyken. "Hybrid integrated mode-locked laser diodes with a silicon nitride extended cavity," Optics Express, vol. 29, no. 10, pp. 15013-15022, May 10, 2021.

85. M. Zhang et al., "Broadband electro-optic frequency comb generation in a lithium niobate microring resonator," Nature, vol. 568, no. 7752, pp. 373-+, Apr 18, 2019.

86. L. Razzari, D. Duchesne, M. Ferrera, R. Morandotti, S. Chu, B. E. Little, and D. J. Moss, "CMOS-compatible integrated optical hyper-parametric oscillator," Nat. Photonics vol. 4, 41-45 (2010).

87. J. S. Levy, A. Gondarenko, M. A. Foster, A. C. Turner-Foster, A. L. Gaeta, and M. Lipson, "CMOS-compatible multiple-wavelength oscillator for on-chip optical interconnects," Nat. Photonics 4, 37-40 (2010).

88. H. Lee, T. Chen, J. Li, K. Y. Yang, S. Jeon, O. Painter, and K. J. Vahala, "Chemically etched ultrahigh-Q wedge-resonator on a silicon chip," Nat. Photonics 6, 369-373 (2012).

89. T. Herr, V. Brasch, J. D. Jost, C. Y. Wang, N. M. Kondratiev, M. L. Gorodetsky, and T. J. Kippenberg, "Temporal solitons in optical microresonators," Nat. Photonics 8, 145-152 (2014).

90. A. G. Griffith, R. K. W. Lau, J. Cardenas, S. Okawachi, A. Mohanty, R. Fain, Y. H. D. Lee, M. Yu, C. T. Phare, C. B. Poitras, A. L. Gaeta, and M. Lipson, "Silicon-chip mid-infrared frequency comb generation," Nat. Commun. 6, 6299 (2015).

91. M. Ferrera, L. Razzari, D. Duchesne and R. Morandotti, Z.Yang, M. Liscidini and J. Sipe, S.Chu, B.E. Little, and D. J. Moss, "Low Power CW Nonlinear Optics in Silica Glass Waveguides", Nature Photonics vol. 2 737 (2008).

92. A. Pasquazi, M. Peccianti, Y. Park, B. E. Little, Sai T. Chu, R. Morandotti, J. Azaña, and D. J. Moss, "Sub-picosecond phase-sensitive optical pulse characterization on a chip", Nature Photonics vol. 5 (10) 618 - 623 (2011).

93. B. J. M. Hausmann, I. Bulu, V. Venkataraman, P. Deotare, and M. Loncar, "Diamond nonlinear photonics," Nature Photonics, vol. 8, no. 5, pp. 369-374, May, 2014.

94. J. Chiles, N. Nader, D. D. Hickstein, S. P. Yu, T. C. Briles, D. Carlson, H. Jung, J. M. Shainline, S. Diddams, S. B. Papp, S. W. Nam, and P. R. Mirin, "Deuterated silicon nitride photonic devices for broadband optical frequency comb generation," Opt. Lett. 43, 1527-1530 (2018).

95. L. Chang, W. Xie, H. Shu, Q. Yang, B. Shen, A Boes, J. D. Peters, W. Jin, C. Xiang, S. Liu, G. Moille, S. Yu, X. Wang, K. Srinivasan, S. B. Papp, K. Vahala, and J. E. Bowers, "Ultra-efficient frequency comb generation in AlGaAs-on-insulator microresonators," Nat. Commun. 11, 8 (2020).

96.  M. Pu, L. Ottaviano, E. Semenova, and K. Yvind, "Efficient frequency comb generation in AlGaAs-on-insulator," Optica 3, 823-826 (2016).

97.  C. Wang, Z. Fang, A. Yi, B. Yang, Z. Wang, L. Zhou, C. Shen, Y. Zhu, Y. Zhou, R. Bao, Z. Li, Y. Chen, K. Huang, J. Zhang, Y. Cheng, and X. Ou, "High-Q microresonators on 4H-silicon-carbide-on-insulator platform for nonlinear photonics," Light: Science & Applications 10, 139 (2021).

98.  H. Jung, S.-P. Yu, D. R. Carlson, T. E. Drake, T. C. Briles, and S. B. Papp, "Tantala Kerr nonlinear integrated photonics," Optica 8, 811-817 (2021).

99.  H. Jung, R. Stoll, X. Guo, D. Fischer, and H. X. Tang, "Green, red, and IR frequency comb line generation from single IR pump in AlN microring resonator," Optica, vol. 1, no. 6, pp. 396-399, Dec 20, 2014.

100. D. J. Wilson, K. Schneider, S. Hönl, M. Anderson, Y. Baumgartner, L. Czornomaz, T. J. Kippenberg, and P. Seidler, "Integrated gallium phosphide nonlinear photonics," Nat. Photonics 14, 57-62 (2020).

101. J. Q. Liu et al., "High-yield, wafer-scale fabrication of ultralow-loss, dispersion-engineered silicon nitride photonic circuits," Nature Communications, vol. 12, no. 1, Apr 16, 2021.

102. Z. C. Ye, K. Twayana, P. A. Andrekson, and V. Torres-Company, "High-Q Si3N4 microresonators based on a subtractive processing for Kerr nonlinear optics," Optics Express, vol. 27, no. 24, pp. 35719-35727, Nov 25, 2019.

103. H. Guo, M. Karpov, E. Lucas, A. Kordts, M. H. Pfeiffer, V. Brasch, G. Lihachev, V. E. Lobanov, M. L. Gorodetsky, and T. J. Kippenberg, "Universal dynamics and deterministic switching of dissipative Kerr solitons in optical microresonators," Nature Physics 13, 94-102 (2017).

104. W. Weng, R. Bouchand, E. Lucas, E. Obrzud, T. Herr, and T. J. Kippenberg, "Heteronuclear soliton molecules in optical microresonators," Nat. Commun. 11, 2402 (2020).

105. M. Yu, J. K. Jang, Y. Okawachi, A. G. Griffith, K. Luke, S. A. Miller, X. Ji, M. Lipson, and A. L. Gaeta, "Breather soliton dynamics in microresonators," Nat. Commun. 8, 14569 (2017).

106. E. Lucas, M. Karpov, H. Guo, M. L. Gorodetsky, and T. J. Kippenberg, "Breathing dissipative solitons in optical microresonators," Nat. Commun. 8, 736 (2017).

107. S. Wan, R. Niu, Z.-Y. Wang, J.-L. Peng, M. Li, J. Li, G.-C. Guo, C.-L. Zou, and C.-H. Dong, "Frequency stabilization and tuning of breathing solitons in Si 3 N 4 microresonators," Photonics Res. 8, 1342-1349 (2020).

108. M. Liu, H. Huang, Z. Lu, Y. Wang, Y. Cai, and W. Zhao, "Dynamics of dark breathers and Raman-Kerr frequency combs influenced by high-order dispersion," Opt. Express 29, 18095-18107 (2021).

109. D. C. Cole, E. S. Lamb, P. Del'Haye, S. A. Diddams, and S. B. Papp, "Soliton crystals in Kerr resonators," Nat. Photonics 11, 671-676 (2017).

110. Y. He, J. Ling, M. Li, and Q. Lin, "Perfect Soliton Crystals on Demand," Laser Photon. Rev. 14, 6 (2020).

111. Z. Lu, H. Chen, W. Wang, L. Yao, Y. Wang, Y. Yu, B. E. Little, S. T. Chu, Q. Gong, W. Zhao, X. Yi, Y. Xiao, and W. Zhang, "Synthesized soliton crystals," Nat. Commun. 12, 3179 (2021).

112. M. Karpov, M. H. P. Pfeiffer, H. Guo, W. Weng, J. Liu, and T. J. Kippenberg, "Dynamics of soliton crystals in optical microresonators," Nature Physics 15, 1071-1077 (2019).

113. X. Xue, Y. Xuan, Y. Liu, P.-H. Wang, S. Chen, J. Wang, D. E. Leaird, M. Qi, and A. M. Weiner, "Mode-locked dark pulse Kerr combs in normal-dispersion microresonators," Nat. Photonics 9, 594-600 (2015).

114. Q.-F. Yang, X. Yi, K. Y. Yang, and K. Vahala, "Stokes solitons in optical microcavities," Nature Physics 13, 53-57 (2017).

115. G. P. Lin, S. Diallo, J. M. Dudley, and K. Chembo, "Universal nonlinear scattering in ultra-high Q whispering gallery-mode resonators," Optics Express, vol. 24, no. 13, pp. 14880-14894, Jun 27, 2016.

116. Y. Bai, M. Zhang, Q. Shi, S. Ding, Y. Qin, Z. Xie, X. Jiang, and M. Xiao, "Brillouin-Kerr Soliton Frequency Combs in an Optical Microresonator," Phys. Rev. Lett. 126, 063901 (2021).

117. Y. He et al., "Self-starting bi-chromatic LiNbO3 soliton microcomb," Optica, vol. 6, no. 9, pp. 1138-1144, Sep 20, 2019.

118. V. Brasch et al., "Photonic chip-based optical frequency comb using soliton Cherenkov radiation," Science 351(6271), 357–360 (2016).

119. C. Joshi, J. K. Jang, K. Luke, X. Ji, S. A. Miller, A. Klenner, Y. Okawachi, M. Lipson, and A. L. Gaeta, "Thermally controlled comb generation and soliton modelocking in microresonators," Opt. Lett. 41, 2565-2568 (2016).

120. S. Zhang, J. M. Silver, L. Del Bino, F. Copie, M. T. M. Woodley, G. N. Ghalanos, A. Ø. Svela, N. Moroney, and P. Del'Haye, "Sub-milliwatt-level microresonator solitons with extended access range using an auxiliary laser," Optica 6, 206-212 (2019).

121. H. Zhou, Y. Geng, W. Cui, S.-W. Huang, Q. Zhou, K. Qiu, and C. Wei Wong, "Soliton bursts and deterministic dissipative Kerr soliton generation in auxiliary-assisted microcavities," Light: Science & Applications 8, 50 (2019).

122. D. C. Cole et al., "Kerr-microresonator solitons from a chirped background," Optica 5, 1304–1310 (2018). (2020).

123. E. Obrzud, S. Lecomte, and T. Herr, "Temporal solitons in micro-resonators driven by optical pulses," Nat. Photonics 11, 600–607 (2017).

124. C. Xiang, J. Liu, J. Guo, L. Chang, R. N. Wang, W. Weng, J. Peters, W. Xie, Z. Zhang, J. Riemensberger, J. Selvidge, T. J. Kippenberg, and J. E. Bowers, "Laser soliton microcombs heterogeneously integrated on silicon," Science 373, 99-103 (2021).

125. B. Shen, L. Chang, J. Liu, H. Wang, Q. Yang, C. Xiang, R. Wang, J. He, T. Liu, W. Xie, J. Guo, D. Kinghorn, L. Wu, Q. Ji, T. J. Kippenberg, K. Vahala, and J. E. Bowers, "Integrated turnkey soliton microcombs," Nature 582, 365-369 (2020).

126. B. Stern, X. Ji, Y. Okawachi, A. L. Gaeta, and M. Lipson, "Battery-operated integrated frequency comb generator," Nature 562, 401-405 (2018).

127. H. Bao, A. Cooper, M. Rowley, L. Di Lauro, J. S. Totero Gongora, S. T. Chu, B. E. Little, G.-L. Oppo, R. Morandotti, D. J. Moss, B. Wetzel, M. Peccianti, and A. Pasquazi, "Laser cavity-soliton microcombs," Nat. Photonics 13, 384-389 (2019).

128. S. B. Papp, K. Beha, P. Del'Haye, F. Quinlan, H. Lee, K. J. Vahala, and S. A. Diddams, "Microresonator frequency comb optical clock," Optica 1, 10-14 (2014).

129. D. T. Spencer, T. Drake, T. C. Briles, J. Stone, L. C. Sinclair, C. Fredrick, Q. Li, D. Westly, B. R. Ilic, A. Bluestone, N. Volet, T. Komljenovic, L. Chang, S. H. Lee, D. Y. Oh, M.-G. Suh, K. Y. Yang, M. H. P. Pfeiffer, T. J. Kippenberg, E. Norberg, L. Theogarajan, K. Vahala, N. R. Newbury, K. Srinivasan, J. E. Bowers, S. A. Diddams, and S. B. Papp, "An optical-frequency synthesizer using integrated photonics," Nature 557, 81-85 (2018).

130. P. Marin-Palomo, J. N. Kemal, M. Karpov, A. Kordts, J. Pfeifle, M. H. Pfeiffer, P. Trocha, S. Wolf, V. Brasch, and M. H. Anderson, "Microresonator-based solitons for massively parallel coherent optical communications," Nature 546, 274-279 (2017).

131. A. Fulop, M. Mazur, A. Lorences-Riesgo, O. B. Helgason, P. Wang, Y. Xuan, D. E. Leaird, M. Qi, P. A. Andrekson, A. M. Weiner, and V. Torres-Company, "High-order coherent communications using mode-locked dark-pulse Kerr combs from microresonators," Nat. Commun. 9 (2018).

132. J. Pfeifle, V. Brasch, M. Lauermann, Y. Yu, D. Wegner, T. Herr, K. Hartinger, P. Schindler, J. Li, and D. Hillerkuss, "Coherent terabit communications with microresonator Kerr frequency combs," Nat. Photonics 8, 375-380 (2014).

133. B. Corcoran, M. Tan, X. Xu, A. Boes, J. Wu, T. G. Nguyen, S. T. Chu, B. E. Little, R. Morandotti, and A. Mitchell, "Ultra-dense optical data transmission over standard fibre with a single chip source," Nat. Commun. 11, 1-7 (2020).

134. Y. Geng, H. Zhou, X. Han, W. Cui, Q. Zhang, B. Liu, G. Deng, Q. Zhou, and K. Qiu, "Coherent optical communications using coherence-cloned Kerr soliton microcombs," Nat. Commun. 13, 1070 (2022).

135. A. Dutt, C. Joshi, X. Ji, J. Cardenas, Y. Okawachi, K. Luke, A. L. Gaeta, and M. Lipson, "On-chip dual-comb source for spectroscopy," Sci. Adv. 4, 9 (2018).

136. C. Bao, Z. Yuan, H. Wang, L. Wu, B. Shen, K. Sung, S. Leifer, Q. Lin, and K. Vahala, "Interleaved difference-frequency generation for microcomb spectral densification in the mid-infrared," Optica 7, 309-315 (2020).

137. P. Trocha, M. Karpov, D. Ganin, M. H. P. Pfeiffer, A. Kordts, S. Wolf, J. Krockenberger, P. Marin-Palomo, C. Weimann, S. Randel, W. Freude, T. J. Kippenberg, and C. Koos, "Ultrafast optical ranging using microresonator soliton frequency combs," Science 359, 887-891 (2018).

138. M.-G. Suh, and K. J. Vahala, "Soliton microcomb range measurement," Science 359, 884 (2018).

139. J. Riemensberger, A. Lukashchuk, M. Karpov, W. Weng, E. Lucas, J. Liu, and T. J. Kippenberg, "Massively parallel coherent laser ranging using a soliton microcomb," Nature 581, 164-+ (2020).

140. E. Obrzud, M. Rainer, A. Harutyunyan, M. H. Anderson, J. Liu, M. Geiselmann, B. Chazelas, S. Kundermann, S. Lecomte, M. Cecconi, A. Ghedina, E. Molinari, F. Pepe, F. Wildi, F. Bouchy, T. J. Kippenberg, and T. Herr, "A microphotonic astrocomb," Nat. Photonics 13, 31-35 (2019).

141. M.-G. Suh, X. Yi, Y.-H. Lai, S. Leifer, I. S. Grudinin, G. Vasisht, E. C. Martin, M. P. Fitzgerald, G. Doppmann, J. Wang, D. Mawet, S. B. Papp, S. A. Diddams, C. Beichman, and K. Vahala, "Searching for exoplanets using a microresonator astrocomb," Nat. Photonics 13, 25-30 (2019).

142. M. G. Suh, Q. Yang, K. Yang, X. Yi, and K. J. Vahala, "Microresonator soliton dual-comb spectroscopy," Science 354, 600-603 (2016).

143. Q.-F. Yang, B. Shen, H. Wang, M. Tran, Z. Zhang, Y. Yang, L. Wu, C. Bao, J. Bowers, A. Yariv, and K. Vahala, "Vernier spectrometer using counterpropagating soliton microcombs," Science 363, 965-968 (2019).

144. M. Yu, Y. Okawachi, A. G. Griffith, N. Picque, M. Lipson, and A. L. Gaeta, "Silicon-chip-based mid-infrared dual-comb spectroscopy," Nat. Commun. 9, 6 (2018).

145. J. Liu, E. Lucas, A. S. Raja, J. He, J. Riemensberger, R. N. Wang, M. Karpov, H. Guo, R. Bouchand, and T. J. Kippenberg, "Photonic microwave generation in the X- and K-band using integrated soliton microcombs," Nat. Photonics 14, 486-491 (2020).

146. X. Xu, J. Wu, M. Tan, T. G. Nguyen, S. T. Chu, B. E. Little, R. Morandotti, A. Mitchell, and D. J. Moss, "Broadband Microwave Frequency Conversion Based on an Integrated Optical Micro-Comb Source," Journal of Lightwave Technology 38, 332-338 (2020).

147. P. Del'Haye, A. Coillet, T. Fortier, K. Beha, D. C. Cole, K. Y. Yang, H. Lee, K. J. Vahala, S. B. Papp, and S. A. Diddams, "Phase-coherent microwave-to-optical link with a self-referenced microcomb," Nat. Photonics 10, 516-520 (2016).

148. X. Xu, M. Tan, J. Wu, R. Morandotti, A. Mitchell, and D. J. Moss, "Microcomb-Based Photonic RF Signal Processing," Ieee Photonics Technology Letters 31, 1854-1857 (2019).

149. X. Xu, M. Tan, J. Wu, T. G. Nguyen, S. T. Chu, B. E. Little, R. Morandotti, A. Mitchell, and D. J. Moss, "Advanced adaptive photonic RF filters based on an optical micro-comb source with 80 taps", Journal of Lightwave Technology 37 (4) 1288-1295 (2019).

150. J. Wu, X. Xu, T. G. Nguyen, S. T. Chu, B. E. Little, R. Morandotti, A. Mitchell, and D. J. Moss, "RF photonics: An optical micro-combs' perspective", IEEE Journal of Selected Topics in Quantum Electronics (JSTQE) 24 (4), 1-20, Article: 6101020 (2018).

151. X. Xu, J. Wu, M. Shoeiby, T. G. Nguyen, S. T. Chu, B. E. Little, R. Morandotti, A. Mitchell, and D. J. Moss, "Advanced RF and microwave functions based on an integrated optical frequency comb source", Optics Express 26 (3), 2569-2583 (2018).

152. M. Tan, X. Xu, B. Corcoran, J. Wu, A. Boes, T. G. Nguyen, S. T. Chu, B. E. Little, R. Morandotti, A. Mitchell, and D. J. Moss, "Broadband microwave and RF photonic fractional Hilbert transformer based on a 50GHz integrated Kerr micro-comb", Journal of Lightwave Technology 37 (24) 6097 – 6104 (2019).

153. M. Tan, X. Xu, B. Corcoran, J. Wu, A. Boes, T. G. Nguyen, S. T. Chu, B. E. Little, R. Morandotti, A. Mitchell, and D. J. Moss, "RF and Microwave Fractional Differentiator based on Photonics", IEEE Transactions on Circuits and Systems: Express Briefs, Volume: 67, Issue: 11, Page(s): 2767-2771 (2020).

154. M. Tan, X. Xu, J. Wu, R. Morandotti, A. Mitchell, and D. J. Moss, "Photonic RF and microwave filters based on 49GHz and 200GHz Kerr microcombs", Optics Communications, 465, Article: 125563 (2020).

155. J. Liu, H. Tian, E. Lucas, A. S. Raja, G. Lihachev, R. N. Wang, J. He, T. Liu, M. H. Anderson, W. Weng, S. A. Bhave, and T. J. Kippenberg, "Monolithic piezoelectric control of soliton microcombs," Nature 583, 385-390 (2020).

156. S. Zhang, J. M. Silver, T. Bi, and P. Del'Haye, "Spectral extension and synchronization of microcombs in a single microresonator," Nat. Commun. 11, 6384 (2020).

157. J. K. Jang, A. Klenner, X. Ji, Y. Okawachi, M. Lipson, and A. L. Gaeta, "Synchronization of coupled optical microresonators," Nat. Photonics 12, 688-693 (2018).

158. T. Yan et al., "All-optical graph representation learning using integrated diffractive photonic computing units." Science Advances 8, 24 (2022).

159. W. Shi et al., "LOEN: Lensless opto-electronic neural network empowered machine vision." Light Sci Appl 11, 121 (2022).

160. F. Ashtiani, A. J. Geers, and F. Aflatouni, "An on-chip photonic deep neural network for image classification." Nature 606, 501–506 (2022).